# A Decentralized Approach for Service Discovery & Availability in P-Grids

Rohit Vashishtha, Ankit Gupta, Piyush Gupta, Shakti Mishra, D S Kushwaha

**Abstract**— The widespread emergence of the Internet as a platform for electronic data distribution and the advent of structured information have revolutionized our ability to deliver information to any corner of the world. Although Service Oriented Architecture (SOA) is a paradigm for organizing and utilizing distributed capabilities that may be under the control of different ownership domains and implemented using various technology stacks and every organization may not be geared up for this. To harness the various software / service resources placed on various systems, we have proposed and implemented a model that is able to establish discovery and sharing in load balanced P-grid environment. The experimental results show that the proposed approach has dramatically lowered the network traffic (nearly negligible), while achieving load balancing in P2P grid systems. Our model is able to support discovery and sharing of resources also.

**Index Terms**—Services, P-Grid, Service Oriented Architecture, load balancing,

——————————— ◆ ———————————

## 1 INTRODUCTION

The widespread emergence of the Internet in the mid 1990s as a platform for electronic data distribution and the advent of structured information have revolutionized our ability to deliver information to any corner of the world. While the introduction of Extensible Markup Language (XML) as a structured format was a major enabling factor, the promise offered by SOAP based web services triggered the discovery of architecture patterns that are now known as Service Oriented Architecture. Service Oriented Architecture is an architectural paradigm and discipline that may be used to build infrastructures enabling those needs (consumers) and those with capabilities (providers) to interact via services across disparate domains of technology and ownership. Services act as the core facilitator of electronic data interchanges yet require additional mechanisms in order to function. Service Oriented Architecture (SOA) is a paradigm for organizing and utilizing distributed capabilities that may be under the control of different ownership domains and implemented using various technology stacks. In general, entities (people and organizations) create capabilities to solve or support a solution for the problems they face in the course of their business. One perceived value of SOA is that it provides a powerful framework for matching needs and capabilities and for combining capabilities to address those needs by leveraging other capabilities. One capability may be repurposed across a multitude of needs.SOA is a "view" of architecture that focuses in on services as the action boundaries between the needs and capabilities in a manner conducive to service discovery and repurposing.

Consider an information system scenario that could benefit from a migration to SOA. Within one organization, three separate business processes use the same functionality, each encapsulating it within an application. In this scenario, the login function, the ability to change the user name, and the ability to persist it are common tasks implemented redundantly in all three processes. This is a suboptimal situation because the company has paid to implement the same basic functionality three times. Such scenarios are highly inefficient and introduce maintenance complexity within IT infrastructures. For example, consider an implementation in which the state of a user is not synchronized across all three processes. In this environment, users might have to remember multiple login username/password tokens and manage changes to their profiles in three separate areas. Additionally, if a manager wanted to deny a user access to all three processes, it is likely that three different procedures would be required (one for each of the applications). Corporate IT workers managing such a system would be effectively tripling their work –and spending more for software and hardware systems. In a more efficient scenario, common tasks would be shared across all three processes. This can be implemented by decoupling the functionality from each

————————————————

- *Rohit Vashishtha, Department of Computer Science & Engineering, Motilal National Institute of Technology, Allahabad.*
- *Ankit Gupta, Department of Computer Science & Engineering, Motilal National Institute of Technology, Allahabad.*
- *Piyush Gupta, Department of Computer Science & Engineering, Motilal National Institute of Technology, Allahabad.*
- *Shakti Mishra, Department of Computer Science & Engineering, Motilal National Institute of Technolog, Allahabad.*
- *D.S. Kushwaha, Department of Computer Science & Engineering, Motilal National Institute of Technology, Allahabad.*

process or application and building a standalone authentication and user management application that can be accessed as a service. In such a scenario, the service itself can be repurposed across multiple processes and applications and the company owning it only has to maintain the functionality in one central place. The common service bus is really a virtual environment whereby services are made available to all potential consumers on a fabric. This is typically referred to as an Enterprise Service Bus (ESB) and has a collection of specialized subcomponents including naming and lookup directories, registry-repositories, and service provider interfaces (for connecting capabilities and integrating systems) as well as a standardized collection of standards and protocols to make communications seamless across all connected devices.

A distributed system consists of multiple autonomous computers that communicate through a computer network. The computers interact with each other in order to achieve a common goal. A computer program that runs in a distributed system is called a distributed program, and distributed programming is the process of writing such programs. The system has to tolerate failure in individual computers. Since, the structure of the system (network topology, network latency, number of computers) is not known in advance, the system may consist of different kinds of computers and network links, and the system may change during the execution of a distributed program. Each computer has only a limited, incomplete view of the system. Each computer may know only one part of the input. Distributed systems can be classified on the basis of number of nodes, location of machines and resource sharing. Major architectures are cluster, grids, peer to peer, cloud computing etc. Grid computing is defined as group of node computation which works together in distributed computing. Each node in grid has computer cluster to perform high performance computing through parallel computation. A computer cluster consists of a headnode (master) and some computational nodes (slaves). Headnode is responsible in communicating with the other headnode in grid, managing computation resource, and scheduling computation jobs to slave.

Grid computing and Grid technologies primarily emerged for satisfying the increasing demand of the scientific computing community for more computing power. Geographically distributed computers, linked through Internet in a Grid-like manner, are used to create virtual supercomputers of vast amount of computing capacity able to solve complex problems from eScience in less time than known before [2]. Thus, within the last years we have witnessed how Grid Computing has helped to achieve breakthroughs in meteorology, physics, medicine and other computing-intensive fields. Examples of such large scale applications are known from optimization, Collaborative/eScience Computing and Data-Intensive Computing, to name a few. Grid computing is still in the development stage, and many challenges are to be addressed. Among these, improving its efficiency is a key issue. The question US is: "How to make use of a large number of computers world-wide, ranging from simple laptops, to clusters of computers and supercomputers connected through heterogeneous networks in an efficient, secure and reliable manner?"[2].

Peer-to-Peer (P2P) architectures are intended to allow autonomous peers to interoperate in a decentralized, distributed manner for fulfilling individual and/or common goals. Peers have equivalent capabilities in providing other peers with data and/or services. The P2P paradigm in general offers a prospect of robustness, scalability and availability of large pool of storage and computational resources.

P-Grid (P2P grid) is a next generation peer-to-peer platform for distributed information management beyond mere file-sharing. P-Grid's most important properties are complete decentralization, self-organization, decentralized load balancing, data management functionalities (update), management of dynamic IP addresses and identities efficient search of resources and services. P-Grid is a truly decentralized structured P2P system which does not require central coordination or knowledge. It is based purely on randomized algorithms and local interactions and targeted at environments with low online probabilities of peers. P-Grid differs from other approaches such as Chord, CAN, Pastry, etc. in terms of practical applicability (especially in respect to dynamic network environments), algorithmic foundations (randomized algorithms with probabilistic guarantees), robustness, and flexibility.

For the majority of Grid systems, scheduling is a very important mechanism. In the simplest of cases, scheduling of jobs can be done in a blind way by simply assigning the incoming tasks to the available compatible resources. Nevertheless, it is a lot more profitable to use more advanced and sophisticated schedulers. Moreover, the schedulers would generally be expected to react to the dynamics of the Grid system, typically by evaluating the present load of the resources, and notifying when new resources join or drop from the system. Additionally, schedulers can be organized in a hierarchical form or can be distributed in order to deal with the large scale of the Grid.

At some abstract level schedulers can be categorized in two categories; one is static scheduling and in this case before starting the process scheduler decides on which node of the Grid the process will run but ones the process started it will continue to run on the same node. There is problem with this scheme let when we started the process node was idle but after some time it become overloaded so the process will still be continue to run on a overloaded node so performance will suffer. On the other hand, in case of dynamic scheduling process is expected to run on that node which will provide best efficiency for the complete system.

Let the process start at any node and after some time that node become overloaded while there are still several nodes in this system that have available resources. So in this case the process should be migrated from that overloaded node to the under loaded node. In general, we can define node any entity having some processing power, may be it can a supercomputer, it can be a huge data-storage entity but in this work we are mainly concern with PC machine with their own processing capability, physical memory (Random Access Memory), and communication interface.

## 1.1 Issues

**Service Sharing in P-Grid:**
As in last section we described the service oriented architecture, the Service Oriented Architecture (SOA) mainly focus on Enterprise level but often a simple Desktop user requires a lot of type of services so we are going to present a new idea that is Service sharing in a grid. As initially we have described a grid is a loosely connected systems and our main concern in this paper is for Desktop Computers. We have a grid of Desktop Computers and in this grid each Desktop Computer must have some services .Let Desktop system "A" require service SERV2 and it don't have it but another Desktop system "B" have this service .Desktop system "B" require a service SERV3 and Desktop system "C" have it. So what option we have in first case A will buy SERV2 and B will buy SERV3 but in this fast growing computer science environment how much services any simple Desktop system can afford to buy .So economically to buy every service by each system user is not a good option. Now suppose two different system user U1 and U2, U1 is a big company owner and U1 use any service SERV4 (24*7) or we can say U1 has high use of this service but U2 is a small shop owner who require this service SERV4 only ones in a year and that is the end of financial year. But U1 and U2 are both paying equally for this service SERV4.So in respect of U2 buying this Service is not good at all. So we can say buying each and every service for each Desktop system is not a good idea. The other option is in first example when A will require service SERV2 it will use the service provided by B and B will use service of C. In second example user U2 will use service owned by user U1.So we can say inside the grid if any system require any service which it don't have but another systems have than it will use the service owned by other systems. This is basically we are going to share the service inside our grid environment , but there are some problems which we are given in next section out of which some problems will be discussed in detail in other sections of this paper.

**Discovery of services:**
As initially we have described grid is a huge collection of Desktop systems , to share service our main concern is how any node N1 will find out the service U1. We can have different network topology on logical level which we can use for discovery of services .Our first option is we can have a central directory of services but this option have some problems one of them is to have a big Server to maintain this huge central directory. Second option can be we will work in a complete distributed way but this will full our network with a huge network traffic. Third option is overlay network which contain the best part of first and second option. We have described all these network topology in previous section .We will describe in detail how the service will be discovered in a huge overlay network of Desktop systems forming a grid.

**Availability of Service:**
Availability of services is a main concern in service sharing environment of Grid. Suppose a service is owned by a number of users and those users are offline at any particular moment of time .So at this time we can say service is unavailable to the active users of the grid, but as we initially described we are sharing service in a grid and grid is a huge infrastructure and users are from different geography location , from different admin domain , from different network domain .The service is owned by users of node in a random fashion , so we can say after significant number of users own the service than some of these random users will be available at any time. So in general we don't have a problem of availability of service but even we cannot sure that if a service is a part of grid it can be available 24*7 .So we can conclude that for simple and normal services we can use sharing service facility, but for critical services (those services which should be accomplished in a strict time duration) sharing services in a grid system is not credible enough.

**Load Imbalance:**
As in last section we described that services are distributed in random fashion there is always imbalance in the numbers of consumers and providers for any particular service. At any time a service owned by a small numbers of node can be required by a huge number of nodes , so if this happens than those small group of users who own the service will suffer high load due to large numbers processing they have to perform , huge network traffic they have to manage and so. So we can say the system having maximum number of services , the system which is giving it's best service for the grid is suffering more .In this paper we will describe in detail how to avoid this problem , and make environment in which any system can allow to use its service by other systems without any load imbalance.

**Legal issues:**
There are some legal issues which are in making a service sharing environment, as we all know maximum services are bound by certification policy. Some of these policies allow only the owner or the service buyer to use that service but other authentication type allow a group of users to use the service. So, all the legal issues should be handled before starting to work on a service sharing environment.

**Load Balancing:**

In computer networking, load balancing is a technique to distribute workload evenly across two or more computers, network links, CPUs, hard drives, or other resources, in order to get optimal resource utilization, maximize throughput, minimize response time, and avoid overload. Using multiple components with load balancing, instead of a single component, may increase reliability through redundancy. The load balancing service is usually provided by a dedicated program or hardware device (such as a multilayer switch or a DNS server).

It is commonly used to mediate internal communications in computer clusters, especially high-availability clusters. If the load is more on a server, then the secondary server takes some load while the other is still processing requests .Typically, two or more web serves are employed in a load balancing scheme. In case one of the servers begins to get overloaded, the requests are forwarded to another server. Load balancing brings down the service time by allowing multiple servers to handle the requests. This service time is reduced by using a load balancer to identify which server has the appropriate availability to receive the traffic.

We present a P-Grid based model for discovering and sharing services. The motive of our proposal is to use the processing capacity of idle or under loaded nodes, by discovering and sharing services with only under loaded nodes.

The rest of the paper is organized as follows. In section 2, we have presented our approach including the components detail. Implementation and simulation results are described in section 3. Section 4 includes related work followed by conclusion.

## 2 PROPOSED APPROACH

We consider a Grid network in which nodes are loosely connected to each other via intercommunication network. In Our model we divide all the nodes in different domains. Domains will be form on the basis of hop count such that the hop count of nodes of one domain is less than a threshold value. In each domain we select one of the nodes as a group leader. Group leader in a domain is not a fixed entity and can be selected dynamically on the basis of computing power, Memory resources, Network bandwidth, CPU cycles.

### 2.1 Components used:

### 2.1.1 Local_Domain_Members :
When any node first wants to be a part of a grid, it has to contact to a grid admin. Grid admin decides in which domain the node will lies and supplies the table Local_Domain_Members to the new node which contains the IP addresses of the nodes of the assigned domain and also add the IP address of the new attached node to the list of Local_Domain_Members of other nodes in its domain.

### 2.1.2 external_nodes_list:
As previously defined when any new node wants to be a part of grid , grid admin along with the local_domain_members list, passes the external_nodes_list as shown in table 1. to the new node, that contains the IP address of some of the nodes of each other domains in the grid. Each node updates its list timely.

TABLE 1: EXTERNAL NODE LIST

| Domain Name | IP Address's |
|---|---|
| Domain Name in string Format | 4 Byte address in string format |

### 2.1.3 other_group_leader_list:
It's a dynamic list that contains the IP address of group leaders of other domains. List consists of a Domain name and its corresponding group leader IP address (Table 2).

TABLE 2: OTHER_GROUP_LEADER LIST

| Domain Name | Group Leader IP address |
|---|---|
| Domain Name in string Format | 4 Byte address in string format |

### 2.1.4 under_loaded_local_list:
Each node in the domain contains this list. This list consists of the IP address of the under-loaded nodes of local domain. This list consists of three sections .First section consist of nodes IP address which are under-loaded in both CPU and RAM usage. Second section consist of nodes IP addresses that are under loaded in CPU only and are not present in first section. Similarly third section consists of nodes which are under-loaded in RAM and are not in first section.

### 2.1.5 external_under-loaded_list:
Each Group leader contains this list. This list consists of IP address of under-loaded nodes of other domains. When any node queries to the group leader for a node list of other domains, then group leader randomly selects some of the nodes IP and send it to the queried node.

TABLE 3: EXTERNAL_UNDER-LOADED_LIST

| Domain Name | IP Addresses of under loaded nodes |
|---|---|
| Domain Name in string Format | 4 Byte address in string format |

### 2.1.6 own_services list:
Every node maintains its list of services that are available to itself. This list exchanges with the all other nodes in its domain.

### 2.1.7 local_domain_servies list:
Each node consists of this list that contains all the services available within its domain and corresponding IP address that provides that service. This list can be formed by exchanging the own_service list with in its domain.

TABLE 4: LOCAL_DOMAIN_SERVIES LIST

| Service Name | IP address |
|---|---|
| Service Names provided by a node | 4 Byte address in string format |

## 2.2 Group Leader Selection:

If any new nodes wants to be a part of a grid, it sends a query to all the nodes which are listed in a table Local_Domain_Members of local domain. As we defined, whole grid is divided among the domains, So domain is considered as a local domain for all the nodes in its domain. We have divided grid into a four domains : A ,B,C,D.

When a node in a domain ups, it sends a query to all other nodes in its domain to get the group leader IP address. If any group leader exists before, it replies to the query and new node updates its entry for the group leader IP. Secondly if there is no group leader then new node waits for a limited amount of time and if it does not receives any reply in that time period it sets itself as a group leader.

## 2.3 Group Leader Functions:

Group leader is also a part of a domain so besides from all the normal activities of any node, Group leader performs a number of other functions are as :

- Connect with the group leaders of other domains.
- Reply of group_leader_query.
- Services information exchange with other group leaders.
- Reply of list of under-loaded nodes of other domain.
- Reply of list of Services of other domains.

### 2.3.1 Connect with the group leaders of other domains:

Once the node has been decided as group leader for a particular domain, it sends a query to the nodes in the list external_nodes_list and queries for their domain group leaders. Nodes on receiving this query replies with the IP address of the Group leader of their domain. Sending Group leader then updates its list of other_group_leader_list. By this exchange of information among the nodes, all the group leaders have the information of other group leaders IP thus all the group leaders are logically connected.

### 2.3.2 Reply of group_leader_query:

Decided group leader replies to the query of group leader IP for all the new nodes wants to be connected to the domain.

### 2.3.3 Load information exchange with other group leaders:

As we described before the general working of a node in a grid, group leader is also a member of the local domain, So at anytime group leader contain the under_loaded_local_list, which contains the IP address of all the under loaded nodes in its domain. When change in the status of the load occurs in any node of the domain, then that node will send the updated status to the group leader and group leader updates it under_loaded_local_list, and in turn send the some of the nodes IP from the updated node list to other group leaders of other domains. The selection of the nodes from the list is on the random basis. So at anytime group leader contains the list of under_loaded_local_list and external_under-loaded_list which contains all the under loaded nodes of local domain and some of the under loaded nodes of other domains respectively.

### 2.3.4 Services information exchange with other group leaders:

Each node consists of a table of a services (own_services) that are available to itself. Each node also contains a list of services (local_domain_servies) and corresponding IP address available to all the nodes in the domain. In local domain each node sends timely a own_service list to all other nodes and in turn each node updates its list of local_domain_service. When any node requires a service that is not in own_service it then checks in local_domain_service. If that service not found in both these tables then it queries to its group leader for this service. Group leader forward this query to other group leaders. Other Group leader, in which domain this service is available replies to the query with the IP address of the node having that service to the querying group leader which in turn sends the IP to the initial node which queries for that service. Now query node has a IP address of the node that has service available. Now with the Peer-to-Peer connection it uses its service.

### 2.3.5 Reply of list of under-loaded nodes of other domain:

When a node does not find any under loaded node in its under_loaded_local_list then it sends query to the group leader. Group leader then randomly selects some of the nodes from the external_nodes_list and sends to the querying node. Now queried node make a P2P connection with randomly chosen one of the nodes got from the Group leader and migrate the process.

### 2.3.6 Reply of list of Services of other domains:

When any group leader of other domains queries to a group leader of local domain for availability of any service, then it sends a IP address of the nodes that provide that service in its domain to the queried group leader.

## 2.4 Working Methodology

First of all when any nodes starts, it sends queries in the form of UDP packet to all other nodes in the local domain whose IP addresses are in the Local_Domain_Members for the group leader IP as shown in fig 1. If group leader exist it replies to the node and node stores its IP but if there is no group leader exist, then after waiting for a time interval it sets itself as a group leader of that domain. Extra functionality of group leader is described previously. Each node timely checks its own status and if it changes, sends to all other nodes in its domain via UDP packet as shown in fig 2. Each node on receiving a UDP packet from other nodes updates its list of under_loaded_local_list.

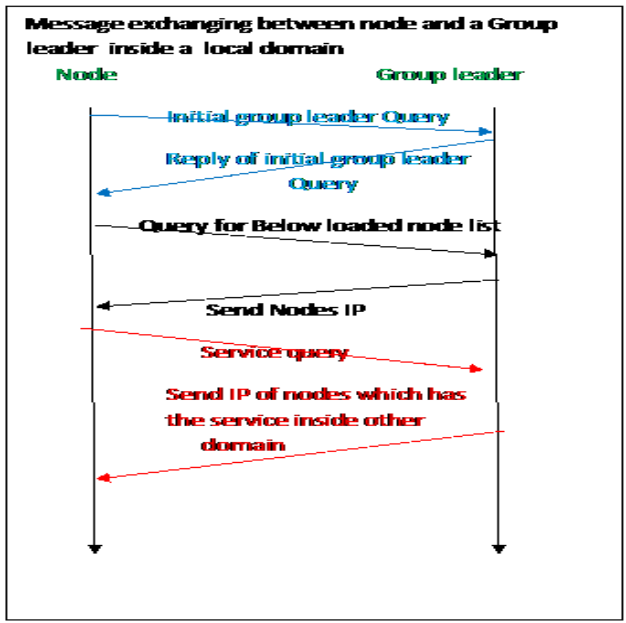

Fig. 1

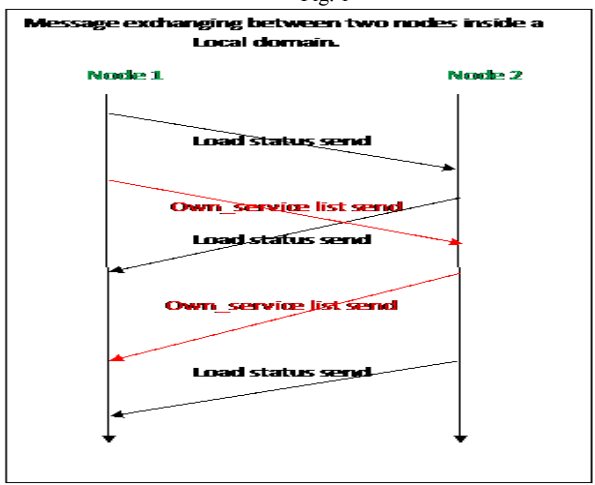

Fig. 2

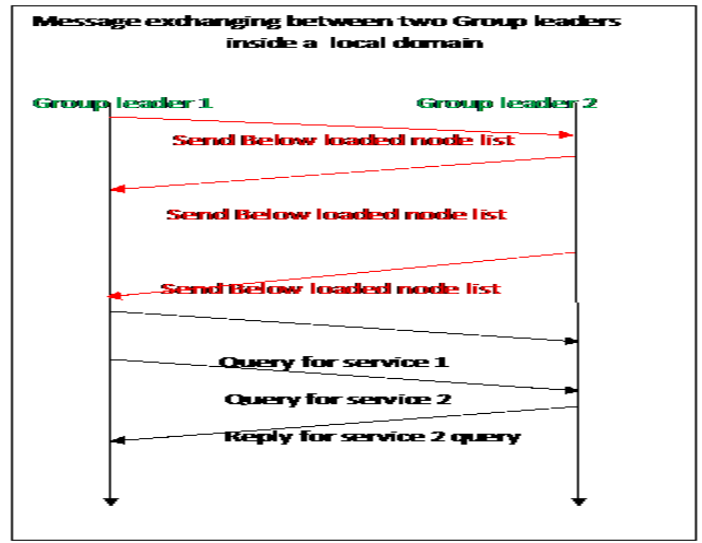

Fig.3

As initially we described group leader also maintains the external_under-loaded_list which can be formed by exchanging under-loaded nodes IP with the other domains group leader. This list can also be updated when under_loaded_local_list of group leader changes as shown in fig 3.

Each node timely checks its own status. Whenever this node becomes overloaded, it has to migrate a process currently running on this node to the other under-loaded node in the grid. So, next work is to decide the node to which process to be migrated. In our model, we first try to migrate a process to any node of local domain than a node of other domain as migrating a process to other domain creates a Network traffic overhead and increase the latency in migration of the process. We select an IP address from under_loaded_local_list and migrate the selected process using MOSIX. If there is any other condition other than defined above, i.e. node is overloaded and there is no suitable node in local domain to whom selected process can be migrated. In this case node sends query to its group leader for nodes of other domain. Group leader replies with it some IP of other domains form external_under-loaded_list .Overloaded node migrates its selected process to randomly selected node from the nodes provided by group leader using MOSIX.

## 3 IMPLEMENTATION AND EXPERIMENTAL STUDY

We have implemented our code in C using Socket programming on Unix based environment, overall functionality of our code is shown through Fig 4. The functionality of each of these threads is defined by algorithms depicted in fig. 5-fig.12.

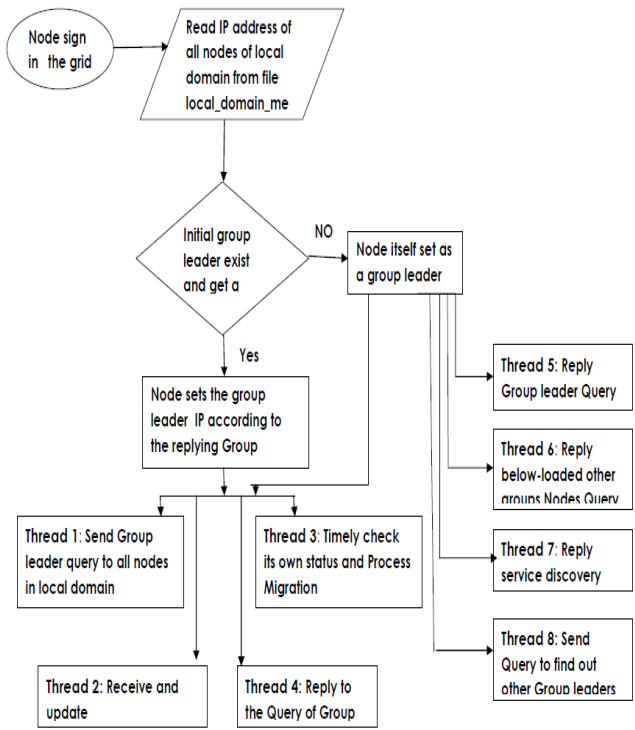

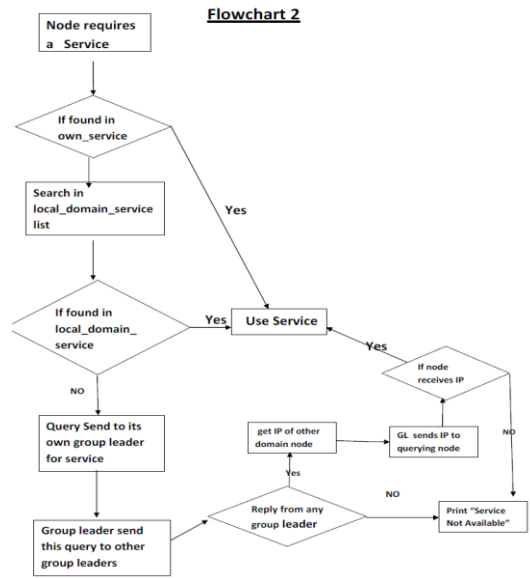

Fig. 4

Fig.13

As we already described that each node has a own_service list and local_domain_servies list. Own_service list consists of all the services that can be offered by a node itself, and by exchanging own_service list inside the local domain through the UDP packets forms the local_domain_servies list. If any node requires any service, then first it looks into its own_service list and if found can be used. If service is not found in this table then it looks into the local_domain_service list. If service is found in this list then it looks for the corresponding IP address of the node that provides that service and forms a direct P2P connection and uses its service. If the required service is not found even in the local_domain_servies list then node queries to the group leader for the IP address of the nodes of other domain the provide that service, group leader in turn queries to the other group leaders of other domains for the IP address of the node providing that service. The procedure described above is shown in fig.13 .

For evaluating the working of our proposed Process Model we Form a sample Grid using 4 Nodes and using MOSIX as a middleware with configuration as:

Node 1. RAM 2GB, Processor Core 2 duo, 2.0 Ghz

IP 172.31.72.42

Node 2. RAM 2GB, Processor Dual core , 1.73Ghz

IP 172.31.72.43

Node 3. RAM 512MB, Processor Core 2 duo,2.0 Ghz

172.31.77.41

Node 4. RAM 1.5GB, Processor Dual core , 1.73Ghz

172.31.77.47

We implemented our proposed algorithm in C using socket programming and Shell Scripting. We divided our grid into 2 domains, Node 1 and Node 2 lies in Domain 1 and Node 3 and Node 4 lies in Domain 2. We observed all the nodes at different time with different load conditions. However, following are the results with time and load variation.

By using two seprate threads nodes are also exchaning the information about the services they have as in following figure initially the node1 has an empty other_service file and node2 has following entries as shown in fig14.

But after some time, exchange of service information occurs and following entries are updated in other_services file of node1 as shown in following fig14.

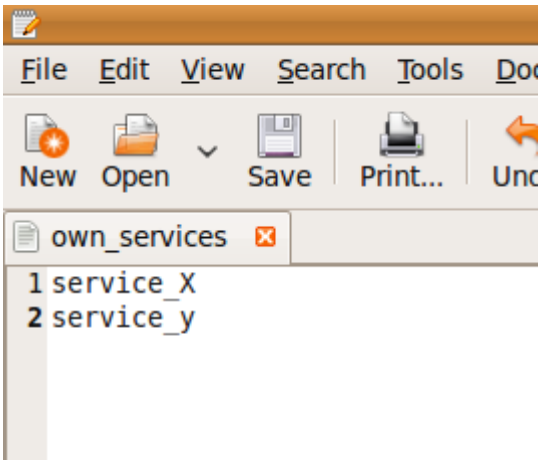

Fig. 14. Own_service list

But after some time ,exchange of service information occurs and following entries are updated in other_services file of node1 as shown in following fig15.

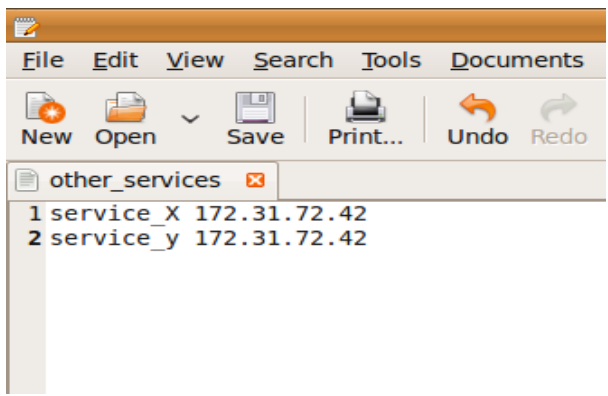

Fig.15 Other_service details with IP address

After some time we made a scenario in which node1 is an normal-loaded and in this condition node2 become overloaded but node2 is not able to find out any node in it's local domain to whom it can migrate it's process but it find node4 in another local domain who is below-loaded and node2 migrate it's process to node4 whose ip is 172.31.72.47.

## 4 RELATED WORK

Subramanian et Al [18] proposed a dynamic Solution to identify a new Web service as a temporary master in runtime when there is any failure in existing master using Fast Bully Algorithm Web Service Community (WSC) is a collection of web service with common functionality. It uses a centralized approach to manage its components i.e. Web services using one Web service has the role of master and the remaining Web services has the role of slave. The work provides a solution for keeping a Web Service Community highly-available to the user or application. It customizes a distributed election algorithm called Fast Bully Algorithm to identify a temporary master Web service when there is any operational failure in existing master Web service of Community. Temporary master Web service is identified from the slave Web services of a community. Although, the major advantage of paper is mechanism for centralized web server failure handling which reduces number of messages, hence low communication cost. But there are still many shortcomings like, sharing of database services with Slave Web services may lead to insecurity since slave services may be unreliable and partial updation /replication leads to redundancy, inconsistency among data files.

The approach in [20] presents an agent based multiple-granularity load balancing middleware model. The basis of the paper is to organize the replicas (host) of same load metric under one agent, thus forming a virtual group. These agents monitor the load for the smaller period of time while the replicas get load information from agents for longer period of time. After deployment, agents register with the load balancer and the reference of agents maps. When new replica is deployed, it gets reference of the agent from the load balancer. But the major limitations with their approach are, number of message increases with agent deployment and the maintenance overhead increases as number of agents increases with number of applications.

Fang. et Al. [19] introduces an infrastructure that parcels out the message-level Web services security processing to load balanced computational Grid nodes while decomposing the processing of web services in to functional components which are running in a distributed fashion, called DEN. It addresses the WS security scalability issue, with the idea derived from pipelining and parallel computer hardware architecture and load balancing mechanisms in distributed systems. However, the routing table maintenance at every modification is an expensive affair. Pipelining of web services may boost the performance and scalability in terms of security however; sequential and interdependent task execution on nodes may lead to poor performance of grid due to various reasons as, I/O waiting, interrupt, signal failure, and deadlock.

## 5 CONCLUSION:

We have been able to establish discovery and sharing in load balanced P-grid environment. The experimental results indicate that, proposed approach has dramatically lowered the network traffic (nearly negligible), while achieving load balancing in P2P grid systems. Our model is able to support discovery and sharing of resources, high availability of services with no specialized utility running on each node that works for homogeneous as well as heterogeneous environment simultaneously.

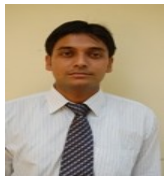

**Rohit Vashishtha** is a students of B.Tech (Information Technology) final year at Motilal nehru National Institute of Technology, Allahabad, India. Their project work is based around High Availability of resources in Distributed Systems

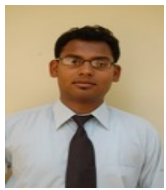

**Ankit Gupta** is a students of B.Tech (Information Technology) final year at Motilal nehru National Institute of Technology, Allahabad, India. Their project work is based around High Availability of resources in Distributed Systems

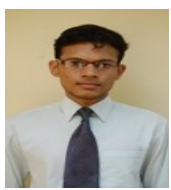

**Piyush Gupta** is a students of B.Tech (Information Technology) final year at Motilal nehru National Institute of Technology, Allahabad, India. Their project work is based around High Availability of resources in Distributed Systems

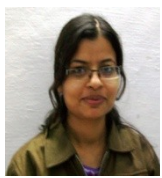

**Ms. Shakti Mishra** received her B.Tech degree from U.P. Technical University, Lucknow, India in the year 2005. She is presently pursuing Ph.D. from Motilal Nehru National Institute of Technology, Allahabad, India under the supervision of Dr. D.S. Kushwaha and Dr. A.K. Misra. She is presently working on load balancing in distributed systems. Her interest area includes Automata theory, Peer to Peer Systems and Mobile Computing.

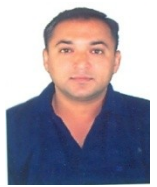

**Dr. D.S.Kushwaha** received his Doctorate Degree in Computer Science & Engineering from Motilal Nehru National Institute of Technology, Allahabad, India in the year 2007 under the guidance of Dr. A.K. Misra. He is presently working with the same Institute as Assistant Professor in the department of Computer Science & Engineering. His research interests include areas in Distributed Systems, Service Oriented Architecture and Data Structures. He has over 30 International publications in various Conferences & Journals.

```
                    Thread 1
    While(1)
    {
       Run script to find local status
       If( Status change from last saved status )
          {
             For (all IP address in local domain )
                {
                   Send New status to all nodes at port 1
                }
          }
     Sleep(1)
    }
```
Fig. 5

```
                    Thread 2
    While(1)
    {
     Always listen at UDP port port 1
     If( receive any load message )
        {
           Update below_loaded_node list according to
                 Senders node IP address
                   }
    }
```
Fig. 6

```
                    Thread 3
While(1)
{
   Node checks its own status
   If (node is overloaded)
    {
        If (below _loaded_node list is not empty)
           Select nodes IP from list randomly
        Else
        {
         Query to group leader for below loaded nodes
             IP of other domain
         Receive reply form group leader as IP list
         Select a nodes IP address from receive nodes
            List from group leader
        }
Migrate selected process to selected node using MOSIX
     } // end of If
Sleep(3);
}
```
Fig. 7

```
                    Thread 4
    While(1)
    {
       Always listen at port 2
       If(receives a query for group leader )
         {
            Reply as a group leader to queried node
         }
    }
```
Fig. 8

```
                    Thread 5
While (1)
{
Always listen at UDP port Port 2
If (receive any query for group leader IP)
   {
        Reply to querying node with group leader IP
   }
}
```
Fig. 9

```
                    Thread 6
  While (1)
  {
  Always listen at UDP port port 3
  If (receive any query for other domains below loaded node)
     {
       Select some nodes IP randomly from
       external_below-loaded_list and reply
       to querying node.
     }
  }
```
Fig. 10

```
                    Thread 7
  While (1)
  {
  Send query to all nodes in external_node_list.
  Listen for reply from replying nodes and update
  group leaders.
  Sleep (100);
  }
```
Fig. 11

```
                    Thread 8
  While (1)
  {
  Always listen at UDP port 4.
    If (receives query for finding any service name)
      {
        Send forwarded query to all group leaders.
        If (receives reply from any group leader)
           Forward this reply to initial query node
      }
  }
```
Fig. 12